\documentclass[letter]{aa} 

\def\ie{i.e.}
\def\eg{e.g.}

\usepackage{natbib}
\usepackage[dvips]{graphicx}

\begin{document}

\title{Sunspot seismic halos generated by fast MHD wave refraction}

\author{E. Khomenko\inst{1,2}, and M. Collados\inst{1}}

\institute{Instituto de Astrof\'{\i}sica de Canarias, 38205,
C/ V\'{\i}a L{\'a}ctea, s/n, Tenerife, Spain;  \email{khomenko@iac.es} \and
Main Astronomical Observatory, NAS, 03680, Kyiv,
Ukraine. \\ }

   \date{Received XXX, 2009; accepted xxx, 2009}

\abstract
   {}  
{We suggest an explanation for the high-frequency power excess
surrounding active regions known as seismic halos.}
{We use numerical simulations of magneto-acoustic wave
propagation in magnetostatic sunspot model.}
{We propose that seismic halos can be
caused by the additional energy injected by high-frequency fast mode waves
refracted in the higher atmosphere due to the rapid increase of
the Alfv\'en speed. Our model qualitatively explains the magnitude
of the halo and allows to make some predictions of its behavior
that can be checked in future observations.}
{}


\authorrunning{Khomenko \& Collados}
\titlerunning{Seismic halos}

   \maketitle


\section{Introduction}

Almost since the discovery of the 5-min solar oscillations
it is well known that the oscillation power is reduced by some
40--60\% in the photospheres of sunspots \citep{Lites+etal1982,
Abdelatif+Lites+Thomas1986, Brown+etal1992, Title+etal1992,
Hindman+Brown1998}. Later it was found that the high-frequency
non-trapped wave power shows a suspicious enhancement in rings
surrounding active regions, both in the photosphere
\citep{Brown+etal1992} and in the chromosphere
\citep{Braun+etal1992, Toner+LaBonte1993}. These power
enhancements are known
as ``halos''. Their observational properties can be summarized as:

(i) The power enhancement is observed at high frequencies between
5.5 and 7.5 mHz for waves that are usually non-trapped in the
non-magnetic quiet Sun.

(ii) The acoustic power measured in halos is larger than in the
nearby quite Sun by about 40-60\% \citep{Hindman+Brown1998,
Braun+Lindsey1999, Donea+etal2000, Jain+Haber2002,
Nagashima+etal2007}.

(iii) The halos are observed at intermediate longitudinal magnetic
fluxes $\langle B \rangle= 50-300$ G, while the acoustic power is
usually reduced at all frequencies at larger fluxes
\citep{Hindman+Brown1998, Thomas+Stanchfield2000, Jain+Haber2002}.

(iv) The radius of the halo increases with height. In the
photosphere the halos are located at the edges of active regions,
while in the chromosphere they extend to a large portion of the
nearby quiet Sun \citep{Brown+etal1992, Braun+etal1992,
Thomas+Stanchfield2000}.

(v) The power increase in the halo is qualitatively similar in
sunspots, pores and plages.

(vi) Significant reflection of the upcoming acoustic radiation at
5--6 mHz is detected in active regions, unlike the behavior of
such high-frequency waves in the quiet Sun
\citep{Braun+Lindsey2000b}.

While several plausible mechanisms have been proposed to explain
the acoustic power reduction for the strongest fields in active
regions \citep[\eg\ MHD mode conversion;][]{Cally+Bogdan1997}, no
accepted theory exists to explain the power enhancement in
acoustic halos. The increase of the high-frequency acoustic
emission, initially proposed by \citet{Brown+etal1992} and
\citet{Braun+etal1992},  seems not to find observational
confirmations, since the observed continuum intensity does not show the
halo effect \citep{Hindman+Brown1998, Jain+Haber2002}.
Alternatively, the latter authors propose that the velocity in the
surroundings of active regions may become field-aligned and some
type of incompressible waves may be responsible for halos. This,
however, lacks any observational evidence.
Recently, \citet{Kuridze+etal2008} suggested yet another mechanism
based on acoustic waves trapped in field-free atmospheres lying
below small-scale magnetic canopies of network cores and active
regions.
Interestingly, halos were observed recently in MHD simulations of
waves in magnetic structures by \citet{Hanasoge2008} and
\citet{Shelyag+etal2009}. Based on his simulations,
\citet{Hanasoge2009} suggests that the power enhancement in halos
is due to magnetic field induced mode mixing resulting in
preferential scattering from low to high wave numbers.

In this Letter we propose a mechanism based on the fast MHD mode
refraction in the vicinity of the transformation layer (where the
Alfv\'en speed $v_A$ is equal to the sound speed $c_S$) that is
capable to explain several observational properties of halos. In
addition, we predict some new properties that can be obtained from
observations in the future to confirm or discard this explanation.

\begin{figure*}
\center
\includegraphics[height=3.1cm]{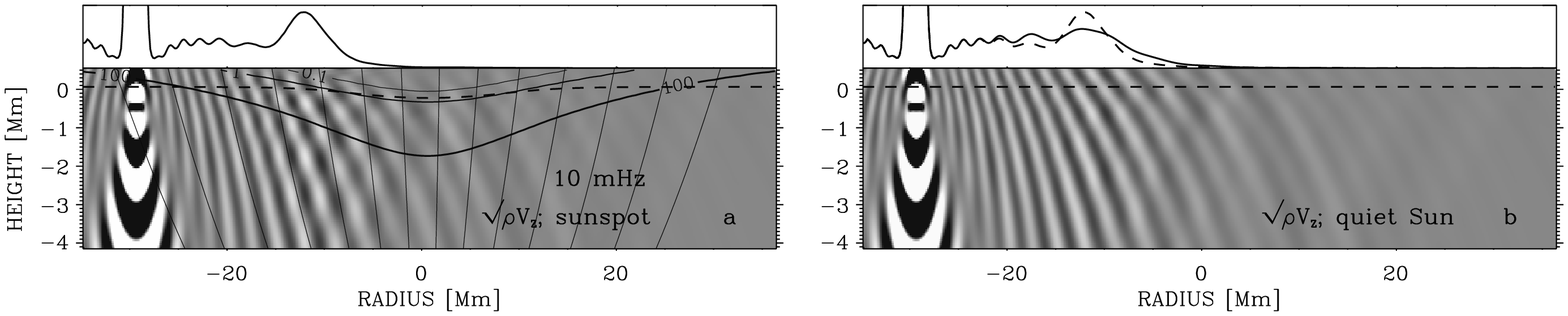}
\includegraphics[height=3.1cm]{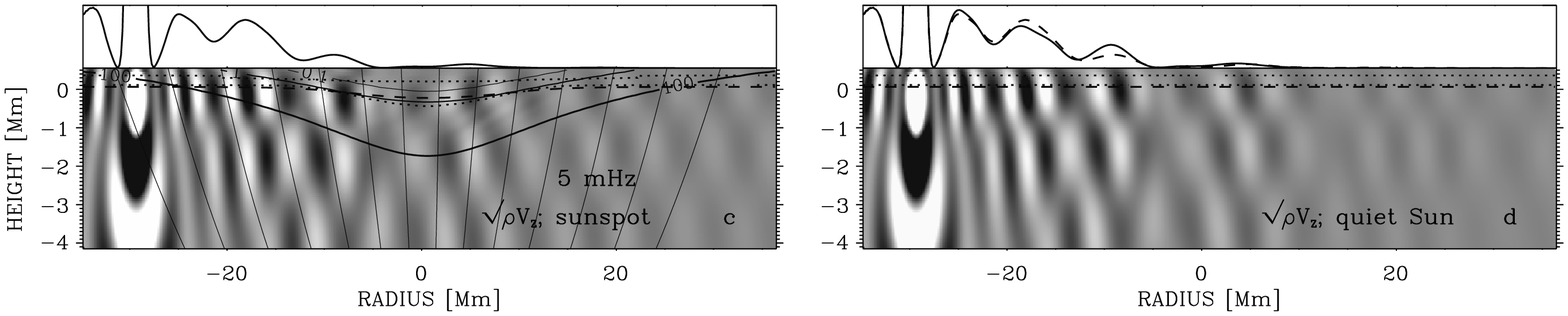}
\includegraphics[height=3.1cm]{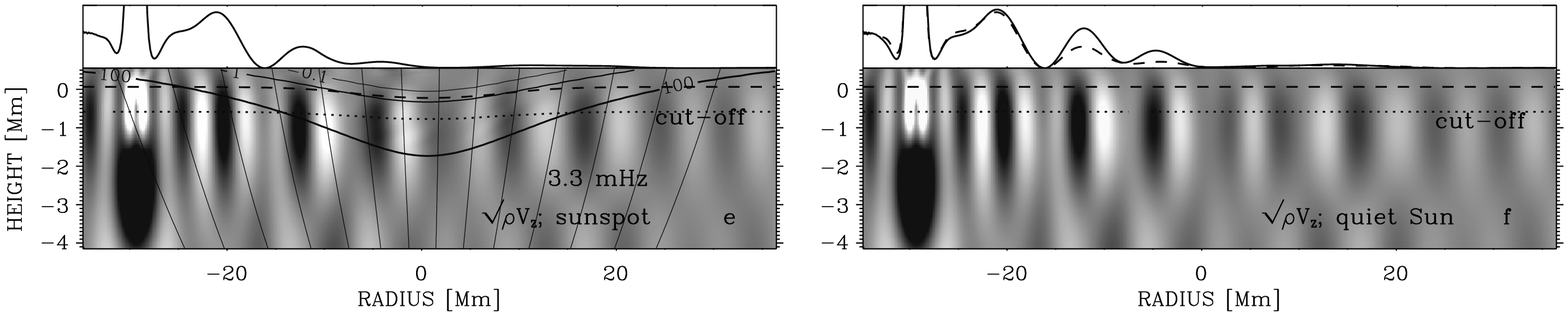}
\caption{Snapshots of the vertical velocity (scaled with the
factor $\sqrt{\rho}$) at elapsed time 60 minutes in the
simulations with harmonic source placed in the sunspot model (left
panels) and in the quiet Sun model (right panels). Panels from top
to bottom: source periods 100, 200 and 300 s. The upper plots on
each panel give the r.m.s. amplitude distribution at the line
formation level. The curve from the sunspot simulations is
repeated in dashed line on the right panels. The contours of
constant $c_S^2/v_A^2$ are marked with numbers. The dashed-line
contours indicate the ``line formation'' heights in both models.
Dotted lines: height where the period of the source coincides with
the cut-off period. The black inclined lines are magnetic field
lines.
}\label{fig:harmonic} \vspace{-0.4cm}
\end{figure*}

\section{Description of methods}

We perform 2D numerical experiments that are essentially similar
to those by \citet{Khomenko+etal2008b}, to study the adiabatic
propagation of magneto-acoustic waves excited by a single source
located at sub-photospheric layers of a magneto-static sunspot
model.  The numerical MHD code is described in detail in
\citet{Khomenko+Collados2006, Khomenko+Collados2007}. The
unperturbed magnetostatic sunspot model is taken from
\citet{Khomenko+Collados2008}. The simulation domain has $80\times
10.5$ Mm$^2$ in horizontal and vertical directions, respectively,
with a resolution of \hbox{$dx=0.15$ Mm} and \hbox{$dz=0.05$ Mm.}
The whole domain contains magnetic field, but it becomes weak and
dynamically unimportant further than $\sim$20 Mm from the sunspot
axis. The maximum field strength in the photosphere is around 1
kG. Our sunspot model has a Wilson depression.
With the help of the SIR radiative transfer code
\citep{RuizCobo+delToroIniesta1992}, from the known distribution
of thermodynamic parameters in geometrical height, we calculated
the optical depth scale, log$\tau_5$, for each horizontal point of
the MHS sunspot model.
The photospheric level defined by the optical depth scale
log$\tau_5=0$ is located 300 km deeper at the sunspot axis
compared to its location 40 Mm far from the axis \citep[see Figs.
1 and 2 in][]{Khomenko+etal2008b}. We define two reference levels
of optical depth: log$\tau_5=0$ (``photosphere'') and
log$\tau_5=-1.6$ (``line formation''). At a horizontal distance
$X=40$ Mm from the sunspot axis, the ``photosphere'' is located
500 km below the top boundary of our simulation domain. The ``line
formation'' level is located 160 km above the ``photosphere''.
At 20 Mm from the axis, the $v_A=c_S$ level is above the ``line
formation'' level by about 300 km, and at the axis it is located
some 200 km below.

In the first set of experiments the source is placed at three
different horizontal distances $X_0=20$, 30 and 35 Mm from the
sunspot axis and at $Z_0=-700$ km below the photosphere. The
temporal behavior of the source is described by a Ricker wavelet
\citep{Parchevsky+Kosovichev2008, Khomenko+etal2008b} with a
central frequency of 3.3 mHz. In the second set of experiments the
source is placed at $X_0=30$ Mm and $Z_0=-450$ km, and it is
harmonic and continuous in time with three different frequencies
$\nu=10$, 5 and 3.3 mHz. In all the cases the sunspot simulations
are accompanied by non-magnetic simulations in the modified model
S of \citet{Christensen-Dalsgaard+etal1996}
\citep[see][]{Parchevsky+Kosovichev2007b} with exactly the same
properties of the source and numerical treatment.  The duration of
simulations is about two physical hours.

\begin{figure}
\center
\includegraphics[width=8cm]{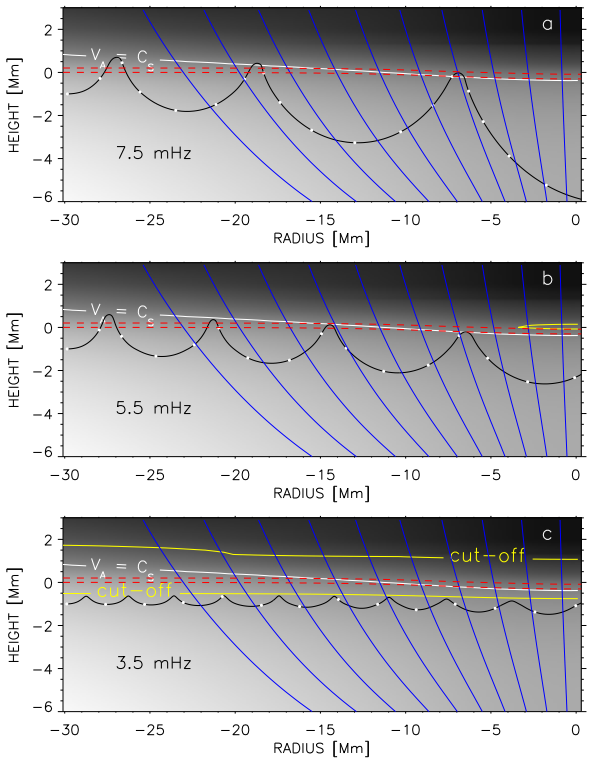}
\caption{Wave paths of the fast mode launched from the lower
turning point at $X=-30$ Mm, $Z=-1$ Mm propagating through the
sunspot model for frequencies 7.5, 5.5 and 3.5 mHz (from top to
bottom). For clearness only the upper part of the model is shown
(not to scale). Blue inclined lines are magnetic field lines. The
yellow contours mark the layer where the wave frequency is equal
to the cut-off frequency. The white contours are $v_A=c_S$. The
background image is log$_{10}(v_A)$. Each white dot on the
trajectory is separated 3 minutes in time. The red dashed lines
are the ``photosphere'' and the ``line formation'' levels.
}\label{fig:wkb} \vspace{-0.4cm}
\end{figure}

\begin{figure*}
\center
\includegraphics[width=14cm]{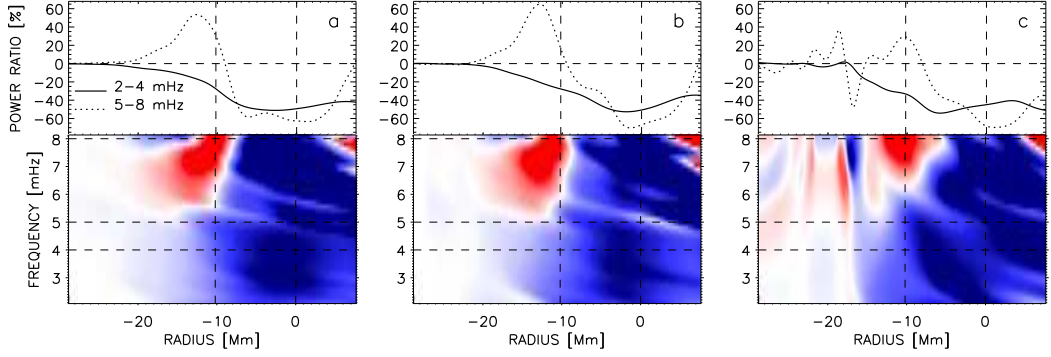}
\caption{Ratio between the photospheric wave power in the sunspot
model relative to the quiet Sun model ($(P_{\rm spot} - P_{\rm
quiet})/P_{\rm quiet}$) as a function of horizontal position and
frequency for three simulations with the wavelet source located at
$X_0=-35$ (a), $-30$ (b) and $-20$ (c) Mm. The color scale at the
bottom panels ranges from -50 to 50\% (red means power excess).
The upper panels show the power ratio averaged over the
low-frequency and high-frequency bands (marked on the figure). The
vertical dashed line located at $X=0$ marks the sunspot axis.
Another dashed line marks the location where $v_A=c_s$ in the
photosphere. }\label{fig:wavelet} \vspace{-0.3cm}
\end{figure*}

\section{Results}

Fig.~\ref{fig:harmonic} shows snapshots of the vertical velocity
in simulations with the harmonic source. When the wave frequency
is  well above the cut-off frequency ($\nu=10$ mHz) the waves are
propagating in the quiet Sun (Fig.~\ref{fig:harmonic}b). This
situation is different in the sunspot model. Significant
reflections can be appreciated in Fig.~\ref{fig:harmonic}a visible
as an interference wave pattern around $X=-20:-10$ Mm. The
reflection of the high-frequency waves is produced in the vicinity
of the transformation $v_A=c_S$ layer. The fast (acoustic-like)
waves generated by the source are transmitted as fast (magnetic)
waves in the upper atmosphere where the Alfv\'en speed is larger
that the sound speed ($v_A> c_S$). In the magnetically dominated
layers these waves are refracted
\citep[see][]{Khomenko+Collados2006} and are returned back to the
sub-photospheric layers where they interfere with the waves coming
from the source. This behavior of waves
is clearly
seen on the movie of this simulation, attached as on-line material
to this paper.
The r.m.s. vertical velocity amplitude measured at the line
formation level (dashed curves in Fig.~\ref{fig:harmonic}a and b)
shows a suspicious bump on the left from the location where the
log$\tau_5=-1.6$ contour crosses the $v_A=c_S$ contour (see upper
plots on each panel). This bump is absent in the quiet Sun
simulation.
We propose that the increase in the high-frequency power in halos
surrounding active regions can be produced by the additional
energy injected by the fast mode waves refracted in the
magnetically dominated layers back to sub-photospheric layers. The
presence of the upward and downward propagating wave energy
manifest itself as the wave interference pattern.
Note that this mechanism does not necessarily imply that the
high-frequency fast waves are trapped in sunspots, since a part of
their energy can leak into the slow mode waves after each mode
transformation at the $v_A=c_S$ layer. However, this mechanism
produces significant reflections of the high-frequency waves that
otherwise are propagating in the quiet Sun.

When the wave frequency is below the photospheric cut-off
frequency ($\nu=3.3$ mHz, Fig.~\ref{fig:harmonic}e, f) the fast
mode waves become evanescent before reaching the transformation
$v_A=c_S$ layer.
In this case neither refraction nor interference can be produced
because of the absence of fast to fast mode transmission.
Fig.~\ref{fig:harmonic}e, f (upper plots) shows that the r.m.s.
amplitude distributions are very similar in the sunspot and the
quiet Sun simulations and that the amplitudes in the sunspot case
are always lower than in the quiet Sun.

The $\nu=5$ mHz case (Fig.~\ref{fig:harmonic}c, d) shows an
intermediate situation, where the waves are reflected both due to
cut-off effects and the magnetic effects. In this case the cut-off
height (dotted lines) almost coincides with the $v_A=c_S$ height.
Some power excess is still present in the sunspot case compared to
the quiet Sun. The bump is weaker and is now located more near the
source as the interference happens at another location due to the
different wavelength. Note that the r.m.s. velocity
distribution is less smooth for the $\nu=5$ and 3.3 mHz compared
to $\nu=10$ mHz, as several nodes are present in the horizontal
(and vertical) direction due to the evanescent character of
waves.

The eikonal solution for the fast mode wave \citep{Cally2006,
Moradi+Cally2008, Khomenko+etal2008b} allows to get a more
complete physical picture of the wave behavior at different
frequencies. Fig.~\ref{fig:wkb} gives the wave paths of the fast
mode waves launched from their lower turning point (that roughly
coincides with the location of the source in the simulations). At
high frequencies ($5.5-7.5$ mHz) the waves penetrate higher up in
the atmosphere above the line formation layer where they are
refracted back down due to the presence of the magnetic field. In
the low-frequency case (3.5 mHz, Fig.~\ref{fig:wkb}c) the waves
are sharply reflected from the cut-off layer and are less affected
by the magnetic field. Note that the power excess (halos) at
high-frequency should form at distances where the refraction of
the fast mode occurs above the line formation layer, \ie\ where
the $v_A=c_S$ layer lies above the photosphere. Otherwise it would
not be detected in spectral observations. This happens in regions
of intermediate field strengths. Under no circumstances can the
halo form in the ``umbral'' zone with strong fields where $v_A$ is
larger than $c_S$ in the photosphere.

A more realistic situation with the source emitting a spectrum of
waves is considered in Fig.~\ref{fig:wavelet}. In this figure we
compare the wavelet source simulations with different source
locations. We represent the ratio between the photospheric Fourier
power in the sunspot and in the quiet Sun at different frequencies
as a function of horizontal distance.
We conclude that, independently of the position of the source, the
overall picture is very similar especially for the sources at
$-35$ and $-30$ Mm. For the source located at the $-20$ Mm from
the axis, the magnetic field plays already some role and modifies
the properties of the source compared to the purely non-magnetic
case, thus contaminating the detailed picture. Still, even in this
case, halo power increase is present around $-10$ Mm, being weaker
than in the other two cases. Fig.~\ref{fig:wavelet} shows a power
excess at high frequencies above 5.5 mHz at distances $-20:-10$ Mm
from the sunspot axis (the latter position coincides with the
location where $v_A=c_S$ in the photosphere). At low frequencies,
a power deficit is observed in the sunspot. The power deficit is
also present at high frequencies in the ``umbral'' region.
The power averaged over both, low-frequency and high-frequency,
bands shows a 40-60\% decrease in the umbra relative to the quiet
Sun. This decrease is because part of the source energy is lost
after the multiple mode transformations
\citep[see][]{Cally+Bogdan1997}. In the high-frequency band,
however, there is an excess of the power up to some 40--50\%. Note
that the magnitude of the halo in our simulations is in a good
agreement with the observed one \citep[\eg ][]{Hindman+Brown1998}.

\section{Discussion and Conclusions}

Our results indicate that the halo effect happens in a natural way
due to additional energy input from the high-frequency fast mode
waves produced after their refraction.
The halo is produced in the photospheric regions where the field
is intermediate implying that the Alfv\'en speed is lower than the
sound speed. The halo is not observed at low frequencies because
these waves are already reflected below the transformation layer.
The halo is not observed in the umbral part of the sunspot because
the refraction happens below the layer visible in spectral line
observations.

In our simulations, the halo is observed at the periphery of the
strong field zone where the field is inclined by some 30--40
degrees. For sunspots with larger field strengths than considered
here, the level $c_S=v_A$ would be located deeper and would
intersect the log$\tau_5=0$ level further from the sunspot axis.
Thus, for models with more intense magnetic field, we expect the
halo will appear at larger distances from the sunspot axis. Also,
by increasing the magnetic field inclination in the penumbra of
the sunspot model, the fast to fast mode transmission will be more
efficient \citep{Cally2006} and we can expect the magnitude of the
halo to be somewhat larger.

Based on our model, we can speculate regarding some observed
properties of the halo as well as others (still undetected) that
may be interesting to observe in the future.

(i) Several observations indicate that the radius of the halo
increases with height
\citep{Brown+etal1992, Braun+etal1992, Thomas+Stanchfield2000}.
This effect can be qualitatively explained by our model. As
follows from \eg\ Fig.\ref{fig:harmonic}, with increasing height,
the $v_A=c_S$ layer is located at progressively larger distances
from the sunspot axis. Observed in chromospheric lines, the
condition necessary to detect the halo (\ie\ that the line
formation layer lies below the $v_A=c_S$ layer) would be fulfilled
at larger distances from the sunspot axis, and, in a natural way,
this would produce halos with larger radius. We can predict that,
after some height in the chromosphere where the whole atmosphere
is magnetically dominated the halos should disappear.

(ii) Our model also explains qualitatively the observations of
\citet{Braun+Lindsey2000b} who detected significant reflection of
the high-frequency waves in active regions.

(iii) The horizontal velocity component (not shown in this paper)
shows a stronger magnitude of the halo effect as the waves
propagate nearly horizontally at the heights where they are
refracted. Thus, we suggest that the magnitude of the halo in
off-center observations should be stronger. We are aware of only
one observations of this kind \citep{Toner+LaBonte1993}, where
apparently no change of the halo magnitude was detected. However,
more observations are required to confirm/discard this conclusion.

(iv) Since the magnitude of the mode transformation and reflection
at the $v_A=c_S$ layer depends on the magnetic field inclination
\citep{Cally2006}, we can speculate that halos, when detected with
high resolution observations,  should show fine structure effects
in active region penumbral filaments, being more pronounced for
horizontal fields.

\acknowledgements   This research has been funded by the Spanish
Ministerio de Educaci{\'o}n y Ciencia through projects
AYA2007-63881 and AYA2007-66502.

\end{document}